\documentclass[fleqn,usenatbib]{mnras}
\usepackage{graphicx}
\usepackage{newtxtext,newtxmath}
\usepackage[T1]{fontenc}
\usepackage{savesym}
\savesymbol{leftrotoavesymbol{uproot}}
\savesymbol{iint}
\savesymbol{iiint}
\savesymbol{iiiint}
\savesymbol{idotsint}
\savesymbol{Bbbk}
\usepackage{amsmath}	
\usepackage{array}
\usepackage{bm}
\usepackage[dvipsnames,svgnames]{xcolor}
\usepackage{soul}
\usepackage[ruled,lined,linesnumbered]{algorithm2e}
\usepackage{amssymb,physics,esint}
\usepackage{multirow}
\usepackage{booktabs}
\usepackage{multirow}
\usepackage{caption}

\DeclareRobustCommand{\VAN}[3]{#2}
\let\VANthebibliography\thebibliography
\def\thebibliography{\DeclareRobustCommand{\VAN}[3]{##3}\VANthebibliography}

\title[Decomposing Gamma-ray Light Curve of PG 1553+113]{Evidence for Magneto-gravitational Processes in Supermassive Black Hole Binary PG 1553+113}

\author[Zhang et al.]{
Haiyun, Zhang$^{1}$
Dahai, Yan$^{2}$\thanks{E-mail: yandahai@ynu.edu.cn}
Li, Zhang$^{2}$\thanks{E-mail: lizhang@ynu.edu.cn}
Niansheng, Tang$^{1}$\thanks{E-mail: nstang@ynu.edu.cn}
\\
$^{1}$Yunnan Key Laboratory of Statistical Modeling and Data Analysis, Yunnan University, Kunming 650091, People's Republic of China\\
$^{2}$Department of Astronomy, Key Laboratory of Astroparticle Physics of Yunnan Province, Yunnan University, Kunming 650091, People's Republic of China
}

\date{Accepted XXX. Received YYY; in original form ZZZ}

\pubyear{2024}

\begin{document}
\label{firstpage}
\pagerange{\pageref{firstpage}--\pageref{lastpage}}
\maketitle

\begin{abstract}

PG 1553+113 has drawn significant attention for its quasi-periodic oscillation (QPO) in $\gamma$-ray variability, though the origin of its variability remains uncertain. 
In this study, we propose a physical mechanism to explain the observed  $\gamma$-ray variability within the framework of a supermassive black hole binary (SMBHB) system, supported by a newly identified component hidden in the light curve.
A detailed analysis for its $\sim$16-year light curve obtained from Fermi-LAT observations is performed by Gaussian process (GP).
As anticipated, the QPO of 2.1 years ($771\pm 8$ days) is effectively captured by the stochastically-driven damped simple harmonic oscillator (SHO) kernel within the under-damped regime, and the overall stochastic nature of the variability is described by the damped random walk (DRW) kernel albeit with an unconstrained damping timescale.
Additionally, our results reveal a previously unrecognized component in active galactic nuclei variability, characterized by the Mat$\acute{\rm e}$rn$-3/2$ kernel, 
which is typically associated with systems undergoing abrupt energy release. 
These findings can be consistently interpreted within the SMBHB framework.
The QPO of $\sim$2.1 years is likely attributed to the orbital motion in a SMBHB system. 
The Mat$\acute{\rm e}$rn$-3/2$ component is interpreted as resulting from magnetic reconnection events triggered by gravitational perturbations of the magnetic field within the jet, occurring as one black hole approaches the other. Meanwhile, in this case, the damping timescale of the common DRW kernel remains unconstrained due to the influence of new perturbations within the system.
\end{abstract} 


\section{Introduction} \label{sec:intro}

PG 1553+113 is one of the brightest high-frequency-peaked BL Lac with a
redshift of $z = 0.49 \pm 0.04$ emitting variable $\gamma$-ray radiation.
It was first reported to have a $\sim 2.2$ yr quasi-periodic oscillation (QPO) in $\gamma$-ray data by Fermi Collaboration \citep{2015ApJ...813L..41A}, supporting by significant cross-correlated variations observed in radio and optical flux light curves.
Subsequently, this QPO was confirmed by the analysis of extended $\gamma$-ray data, 
furthermore it was reported to be consistent with the recurrence of the X-ray primary flare \citep{2018ApJ...854...11T,2021ApJ...922..222H}.
   
The quasi-periodicity observed in the light curves of blazars suggests a physical periodic modulation occurring within the radiating region, though the underlying mechanism remains a topic of debate. Current theories include geometric effects, such as a helical jet structure or jet precession \citep[e.g.,][]{2018MNRAS.478.3199B,2021ApJ...922..222H,2023ApJ...945..146G}, as well as beaming effects \citep[e.g.,][]{1998MNRAS.293L..13V,2018ApJ...867...53Y}. Another possibility is quasi-periodicity arising from processes that feed the jet, such as pulsational instabilities in the accretion flow. Additionally, a supermassive black hole binary (SMBHB) system presents a compelling scenario for explaining QPO phenomena, as discussed in \cite{2015ApJ...813L..41A}.
PG 1553+113 was suggested to be an SMBHB candidate \citep{2017MNRAS.465..161S,2018ApJ...854...11T,2019ApJ...875L..22C,2021ApJ...922..222H}, because of a quasi-periodicity of approximately 2.2 years in its $\gamma$-ray light curve \citep{2018ApJ...854...11T,2020ApJ...895..122C}. In this scenario, the primary black hole’s jet is thought to be periodically perturbed by the gravitation of the secondary black hole, 
inducing jet instabilities that lead to periodic variability in the non-thermal radiation \citep[e.g.,][]{2017ApJ...836..220C}.

By reliably identifying QPOs and examining variability patterns, we can gain insights into the underlying mechanisms driving the behaviors. 
It is actually difficult to accurately identify QPOs.  
The classical analysis technique, e.g., Fourier-like transform methods, may falsely identify QPOs when the size and the quality of sampling data are not well considered \citep{2019MNRAS.482.1270C}.  
An essential aspect of QPO identification and analysis is understanding and modeling the stochastic noise in the data, which significantly impacts the confidence in QPO detection \citep[e.g.,][]{2016MNRAS.461.3145V}. 
Red, or correlated, noise is particularly relevant, as it is common in light curves of active galactic nuclei (AGNs), affecting overall variability modes and the interpretation of variability. High-energy blazar light curves also often contain significant stochastic components (colored noise), posing challenges for period detection.
All of these remind us that the issue of the reality of QPOs are under debate and requires careful consideration \citep[e.g.,][]{2021ApJ...907..105Y,2021ApJ...919...58Z}.

Gaussian Process (GP) modeling offers a powerful probabilistic approach that operates in the time domain to extract detailed information from light curves \citep{2023ARA&A..61..329A}. This non-parametric framework provides flexibility to model complex, unknown functions by selecting suitable kernel (or covariance) functions. The choice of kernel is crucial, as it dictates properties like smoothness and periodicity, thereby enhancing the reliability with which we characterize observed light curves. The GP method has been therefore used to independently verify the reliability of QPO signals \citep[e.g.,][]{2020ApJ...895..122C,2021ApJ...907..105Y,2024MNRAS.531.4181O}.

The first found $\gamma$-ray QPO with high confidence (5$\sigma$) in the non-blazar PKS 0521-36 is an example of QPO identification by GP method. The robustness of this QPO has been examined by different method \citep{2021ApJ...919...58Z}.
\cite{2024MNRAS.531.4181O} has explored and compared the two methods for modeling stochastic and QPO behavior of time series, namely frequency domain analysis based on like-fourier transform and GP method. 
They found that when modeling complex QPO behavior, GP regression results are often more precise, and systematically more accurate in irregular time-sampling regimes.
It should be noted that periodic signals that are not captured by GP methods \citep[e.g.,][]{2021ApJ...907..105Y} are not necessarily false positives; rather, they may be influenced by the chosen model for stochastic noise.

For AGN long-term light curves, the DRW model is commonly used to describe stochastic noise \citep[e.g.,][]{2019ApJ...885...12R,2021Sci...373..789B,2022ApJ...930..157Z,2023ApJ...944..103Z}, and the stochastically driven damped simple harmonic oscillator (SHO) kernel \citep{2017AJ....154..220F,2019PASP..131f3001M} has been used to characterize QPO behaviors \citep{2021ApJ...919...58Z,2021ApJ...907..105Y}.
The SHO kernel has also been used to explore the stochastic variability of blazar's long-term light curves that extend beyond the DRW model \citep[e.g.,][]{2022ApJ...930..157Z}.
In addition to the two models above commonly used in GP methods, Mat$\acute{\rm e}$rn class models can describe stochastic processes but have little application in the field of astronomy.

In this work, we have updated the Fermi light curve of PG 1553+113 through September 20, 2024, and reanalyzed the variability using the GP method with different kernel functions. We apply various basic stochastic models, combined with a periodic model, as kernel functions to characterize the light curve and search for the QPO signal, aiming to better capture both the stochastic noise and periodic components in the $\gamma$-ray light curve of PG 1553+113. This approach allows us to decompose the $\gamma$-ray light curve and gain insights into the underlying physical processes.

\section{Data and Method}\label{sec:method}

The Large Area Telescope on the Fermi Gamma-ray Space Telescope (Fermi-LAT) provides continuous monitoring of the high-energy $\gamma$-ray sky.
This work uses LAT observations of PG 1553+113 covering $\sim$16 years (from
August 4, 2008 to September 20, 2024; MJD 54682.65–60573.5) in the energy range of 0.1-100 GeV.
We get the 7 days binning light curve data directly from Fermi-LAT light curve repository (LCR)\footnote{\url{https://fermi.gsfc.nasa.gov/ssc/data/access/lat/LightCurveRepository/}} which is performed with the standard Fermi-LAT science tools (version v11r5p3) and unbinned likelihood analysis, and 
the spectral model used for PG 1553+113 is LogParabola (LP).
We then excluded time bins with TS$<$25 to construct a new light curve of PG 1553+113.

GP is a powerful class of statistical model to capture the characteristics in the time series \citep[e.g.,][]{2017AJ....154..220F,2023ARA&A..61..329A}.
The choice of kernel function in GP can indeed be guided by prior knowledge about the data.
The Fermi-LAT light curve of PG 1553+113 has been studied 
to be periodic \citep[e.g.,][]{2015ApJ...813L..41A,2018ApJ...867...53Y}.
We use this prior knowledge about the data in the GP analysis.
The SHO kernel \citep{2019PASP..131f3001M}, 
can be naturally used to describe a periodic behavior \citep[e.g.,][]{2017AJ....154..220F,2021ApJ...919...58Z}. 
For irregular fluctuations reflected in time series, 
DRW is a commonly used kernel in AGN variability studies \citep[e.g.,][]{2021Sci...373..789B,2022ApJ...930..157Z}.
Both SHO and DRW kernel have been introduced in detail in our previous works \citep{2021ApJ...919...58Z,2022ApJ...930..157Z,2023ApJ...944..103Z,2024ApJ...971...26T}, and will not be repeated here.

Here we introduce the Mat$\acute{\rm e}$rn class covariance functions \citep{2006gpml.book.....R} in our analysis.
The covariance function is generally parameterized by a smoothness parameter $\nu$, 
which controls the level of smoothness of the function.
It  takes a simplified form when $\nu$ is half-integer, 
$\nu=p+1/2$ where $p$ is a non-negative integer.
In this case, the covariance function becomes a product of an exponential term and a polynomial of order $p$ \citep{2006gpml.book.....R}.
We use an interest case of $\nu=3/2$, 
and the Mat$\acute{\rm e}$rn$-3/2$ function includes an exponential multiplied by a linear polynomial.     
The Mat$\acute{\rm e}$rn$-3/2$ covariance/kernel function is expressed as
\begin{equation}
  k (\tau)=\sigma^{2}(1+\frac{\sqrt{3}\tau}{\rho}){\rm exp}(-\frac{\sqrt{3}\tau}{\rho})\ ,
    \label{maternkernel}
\end{equation}
where $\rho$ is a characteristic timescale in time series, 
$\sigma^{2}$ is the overall variance (also known as amplitude), 
and $\tau$ represent the time lag between two measurements.
The power spectral density (PSD) of Mat$\acute{\rm e}$rn$-3/2$ is flat at low frequencies, while at high frequencies, it decays with a power-law index of $-4$ \citep{2006gpml.book.....R}, 
that is the PSD in the form of broken power-law with the broken frequency at $f_{\rm b}=1/(2\pi\rho)$.
The details on the PSDs of the DRW and SHO kernels are presented in \cite{2024ApJ...971...26T}.
The Mat$\acute{\rm e}$rn$-3/2$, DRW and SHO kernel each have different PSDs, which means that they describe different types of time series characteristics.

We use the combinations of the DRW, Mat$\acute{\rm e}$rn-3/2 and SHO kernel to fitting the $\gamma$-ray light curve of PG 1553+113 over 16 years.
The fitting is performed by 
the {\it celerite} package \citep{2017AJ....154..220F}, 
and the procedure and goodness-of-fit criteria follow the details presented in our previous work \citep{2021ApJ...919...58Z,2022ApJ...930..157Z,2023ApJ...944..103Z}.

\section{Results}\label{sec:results}
\begin{figure*}
    \centering
    {\includegraphics[width=0.8\linewidth]{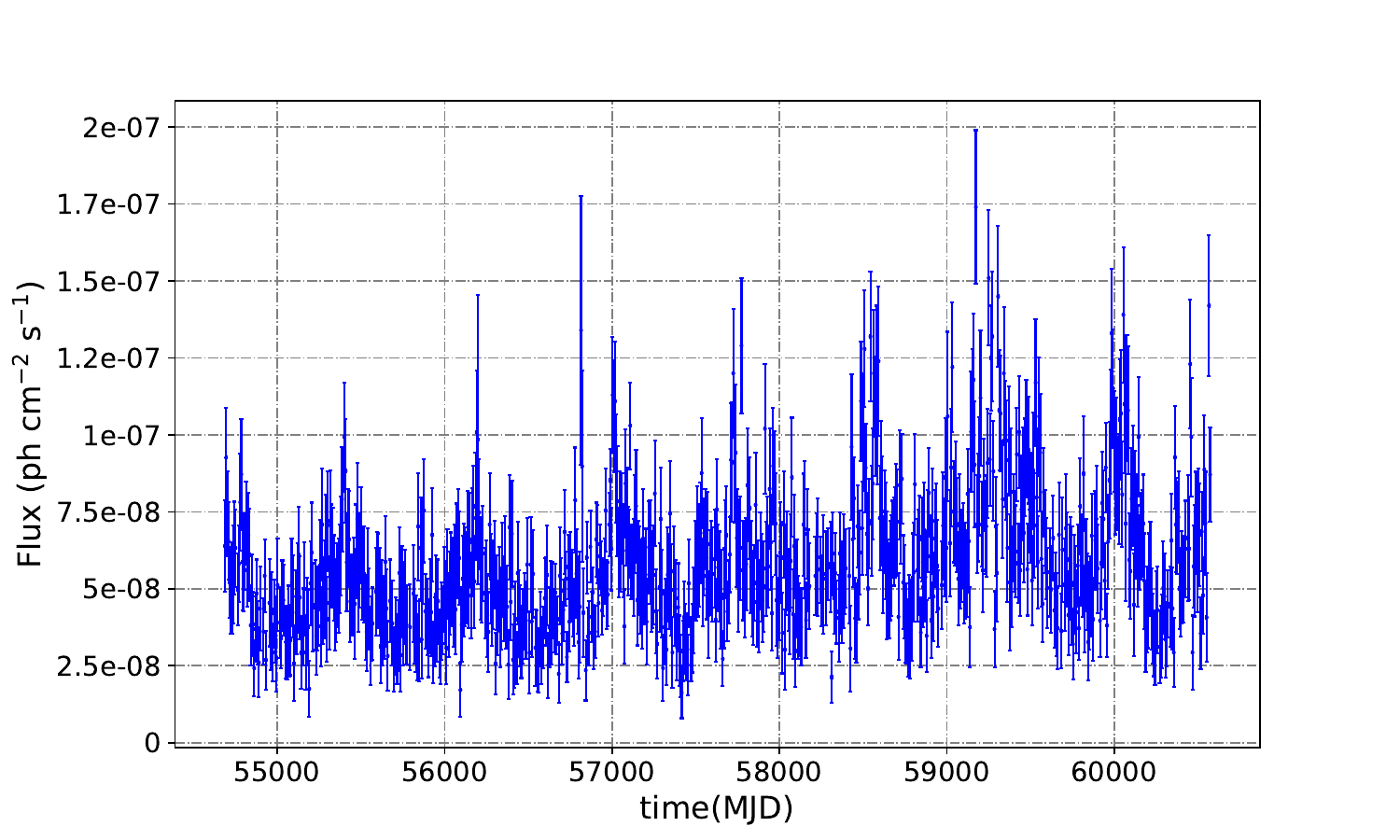}} 
\caption{The Fermi-LAT light curve of PG 1553+113, with a 7-day binning, spans approximately 16 years.}
\label{fig:LC}
\end{figure*}

The 7 days binning $\gamma$-ray light curve of PG 1553+113 is shown in Figure~\ref{fig:LC}.
We first fit the light curve with DRW+SHO and Mat$\acute{\rm e}$rn$-3/2$+SHO models respectively.
Before fitting, we choose the specific kernels and define the priors on the parameters. We adopt a log-uniform density for each parameter as listed in Table~\ref{tab:prior}. 
Note that the prior distribution of $Q$ spans from an extremely small value (the over-damped mode: aperiodic mode) to an extremely large value (high-quality oscillation mode: periodic mode).
This means that we do not have a prior belief about the QPO signature in PG 1553+113.

\begin{figure*}
    \centering
    {\includegraphics[width=0.8\linewidth]{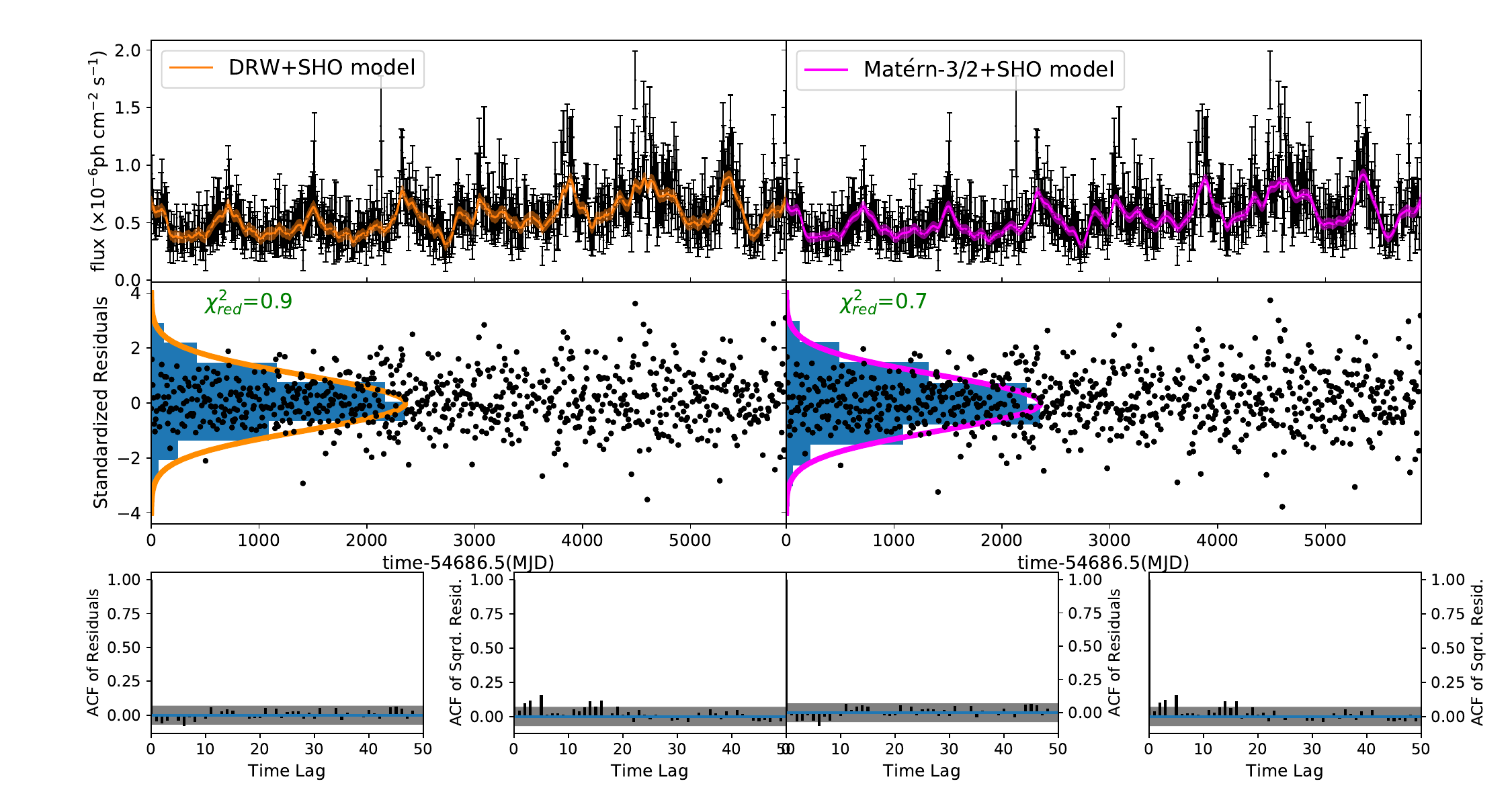}} 
\caption{Fitting results of DRW+SHO (left column) and Mat$\acute{\rm e}$rn$-3/2$+SHO (right column) models for the Fermi-LAT light curve of PG 1553+113. In each column, the top panel presents the observed light curve (black points) and the modeled light curve (orange/magenta line). 
The middle panel shows the standardized residuals (black points) and their distribution, along with the fitted normal distribution (orange/magenta solid line).
In the bottom panel, the ACF and $\rm ACF^{2}$ of the residuals are plotted, with the 95$\%$ confidence limits for white noise shown in gray.
\label{fig:DMSfit}}
\end{figure*}

The fitting results are shown in Figure~\ref{fig:DMSfit}.
There appears to be no discernible distinction between the two combinations, 
according to the goodness-of-fit criteria and the values of the corrected Akaike information criterion ($\rm AIC_{\rm C}$) for DRW+SHO ($\rm AIC_{\rm C}=-590$) and Mat$\acute{\rm e}$rn$-3/2$+SHO ($\rm AIC_{\rm C}=-588$).

\begin{table}
 \caption{Parameters and Priors for Each Model.}
 \resizebox{0.45\textwidth}{!}{
 \begin{tabular}{lll}
 \toprule
 Model & Parameter & Prior  \\
 \midrule
 \multirow{5}{*}{DRW+SHO} & ln($S_{0}$) & Log-uniform(-20, 20) \\
 & ln($Q$) &  Log-uniform(-20, 20) \\
 & ln($\omega_{0}/\rm rad\cdot day^{-1}$) &  Log-uniform(-5, -4) \\
 & ln($\sigma_{\rm DRW}$) &  Log-uniform(-5, 1) \\
 & ln($\tau_{\rm DRW}/\rm day$) &  Log-uniform(2,11) \\
\toprule
\multirow{5}{*}{Mat$\acute{\rm e}$rn$-3/2$+SHO} & ln($S_{0}$) &  Log-uniform(-20, 20) \\
 & ln($Q$) &  Log-uniform(-20, 20) \\
 & ln($\omega_{0}/\rm rad\cdot day^{-1}$) &  Log-uniform(-5, -4) \\
 & ln($\sigma$) &  Log-uniform(-3, 5) \\
 & ln($\rho/\rm day$) &  Log-uniform(2, 6) \\
\midrule
\multirow{7}{*}{Mat$\acute{\rm e}$rn$-3/2$+DRW+SHO} & ln($S_{0}$) &  Log-uniform(-20,20) \\
 & ln($Q$) &  Log-uniform(-20, 20) \\
 & ln($\omega_{0}/\rm rad\cdot day^{-1}$) &  Log-uniform(-5, -4) \\
 & ln($\sigma_{\rm DRW}$) &  Log-uniform(-5, 1) \\
 & ln($\tau_{\rm DRW}/\rm day$) &  Log-uniform(2, 11) \\
 & ln($\sigma$) &  Log-uniform(-3, 5) \\
 & ln($\rho/\rm day$) &  Log-uniform(2, 6) \\
\bottomrule

\end{tabular}
}
\label{tab:prior}

\end{table}

\begin{figure}
    \centering
    {\includegraphics[width=1\linewidth]{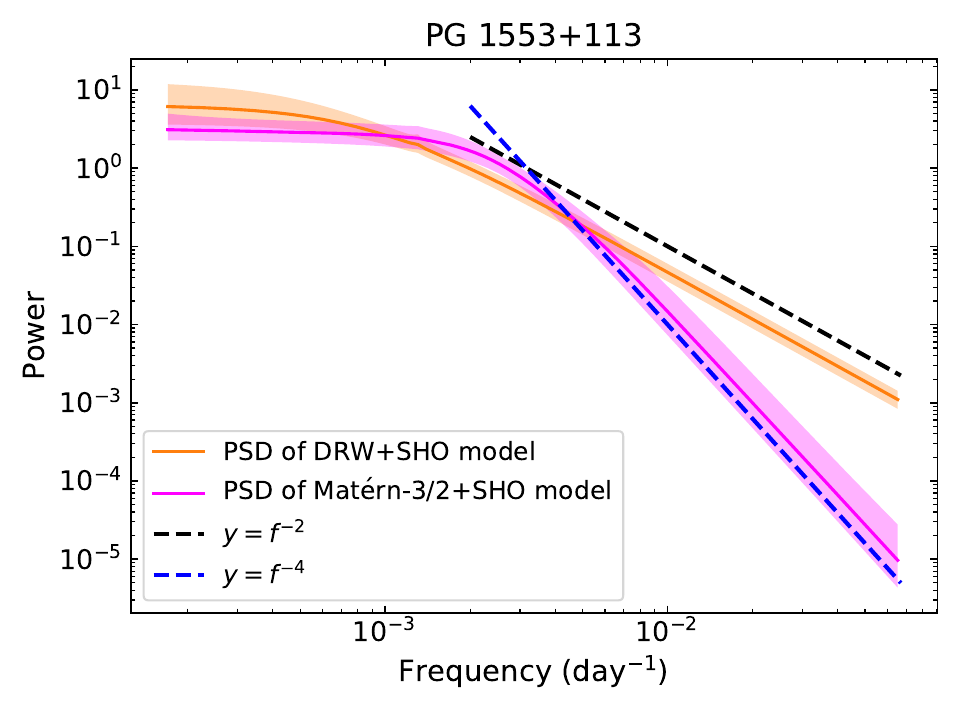}} 
\caption{The PSD constructed from the modeling results with the corresponding colored region indicating the 1$\sigma$ confidence interval.
\label{fig:DMSpsd}}
\end{figure}

\begin{figure*}
    \centering
    {\includegraphics[width=0.49\linewidth]{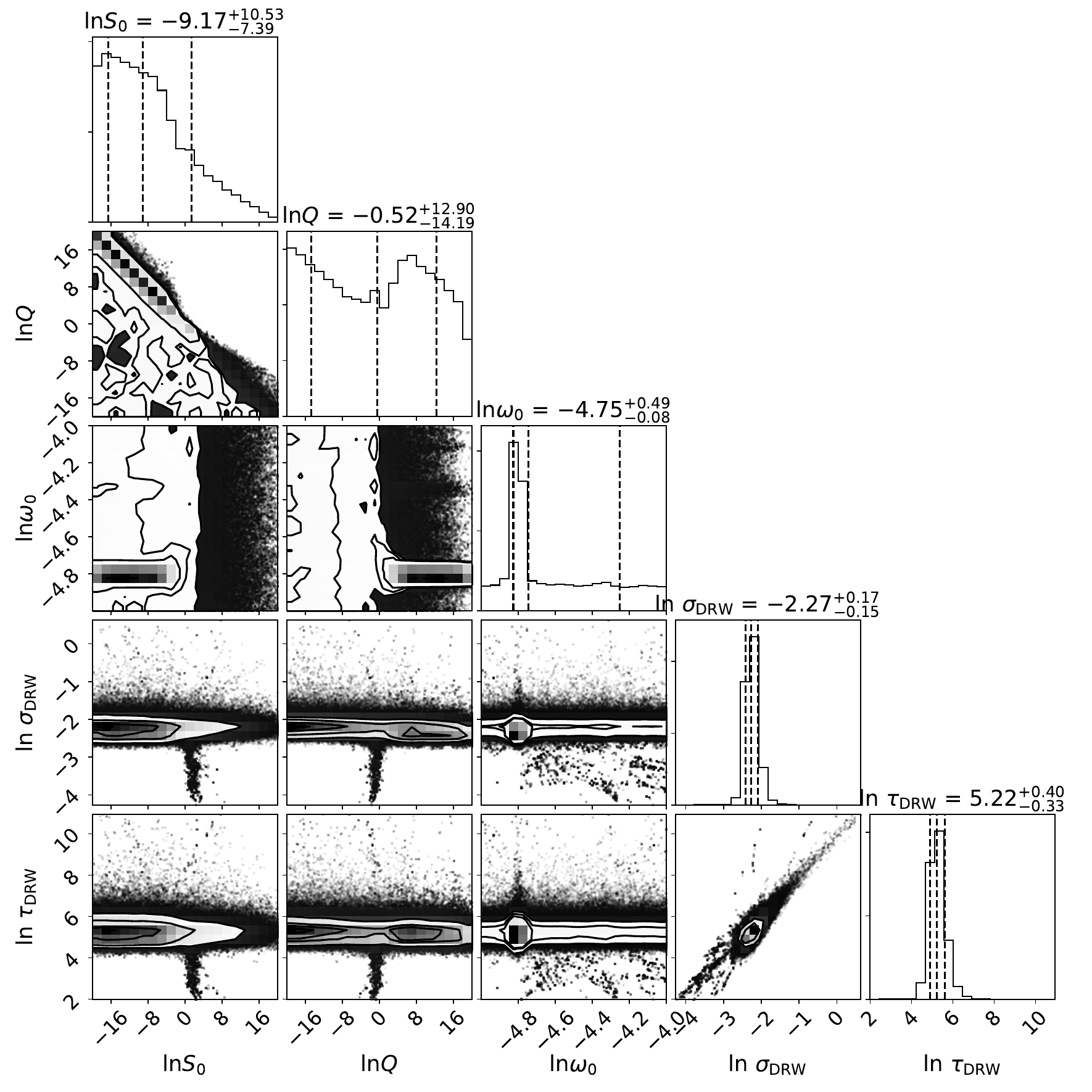}} 
    {\includegraphics[width=0.49\linewidth]{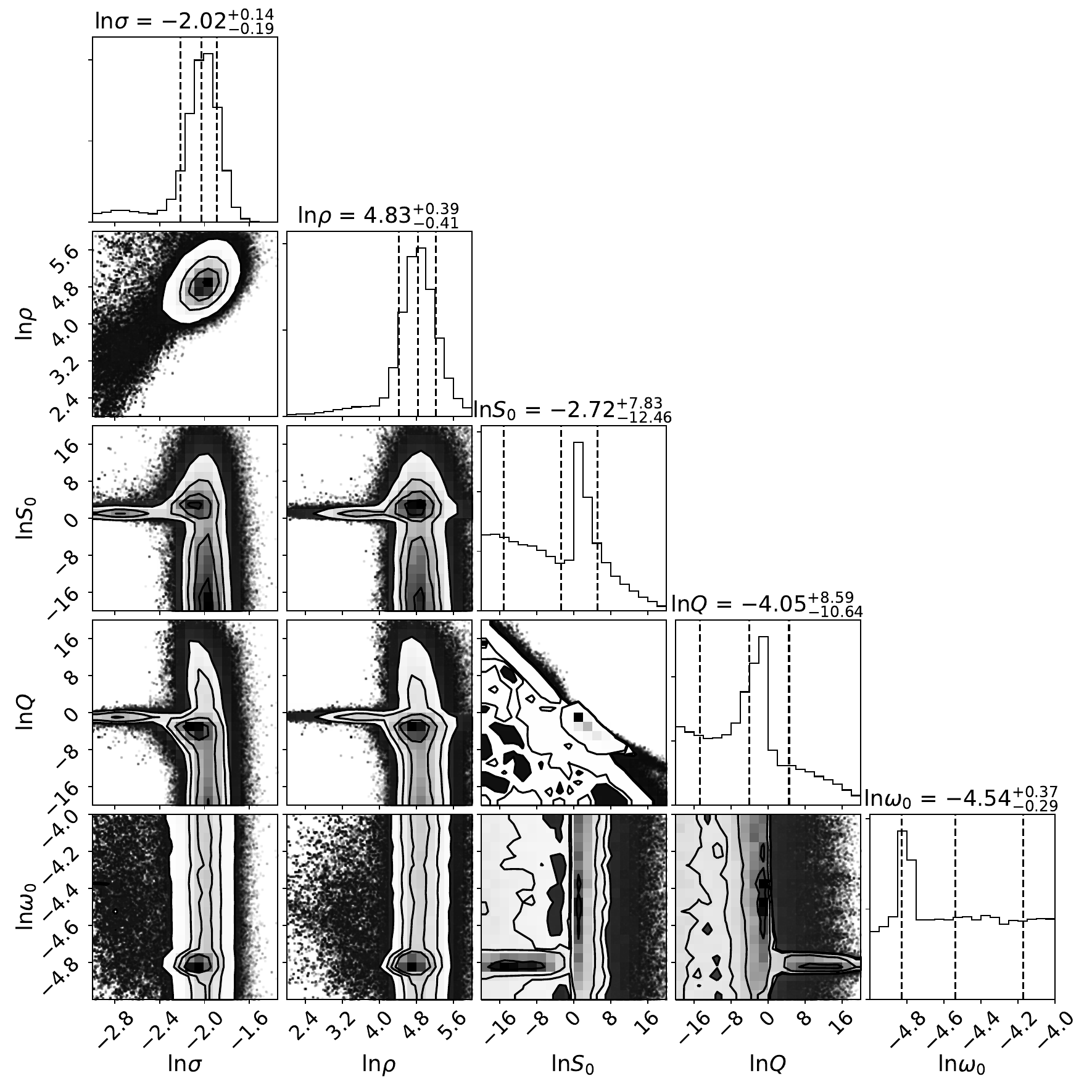}}    
\caption{Posterior distributions of the model parameters. The left panel is for DRW+SHO model, and the right panel is for Mat$\acute{\rm e}$rn$-3/2$+SHO model. The vertical dotted lines mark the median value and 68$\%$ confidence intervals for the parameter distributions.
\label{fig:DMSparams}}
\end{figure*}

Figure~\ref{fig:DMSpsd} gives the resulting PSDs of the two models. 
Each PSD is almost dominated by the stochastic background, i.e., DRW PSD (the orange line) and Mat$\acute{\rm e}$rn$-3/2$ PSD (the magenta line), respectively.
The PSD of SHO is overshadowed by that of the DRW/Mat$\acute{\rm e}$rn$-3/2$.
Indeed, the parameters of SHO in the two models are poorly constrained as shown in Figure~\ref{fig:DMSparams}.
In both models, the $Q$ values span a wide range, covering regimes from the over-damped mode to the under-damped mode (Figure~\ref{fig:DMSparams}). 
The small mean values of $Q$ result in the absence of a QPO signal. Due to the strong degeneracy between $Q$ and $S_{\rm 0}$, the $S_{\rm 0}$ also remains unconstrained.

\begin{figure}
    \centering
    {\includegraphics[width=1\linewidth]{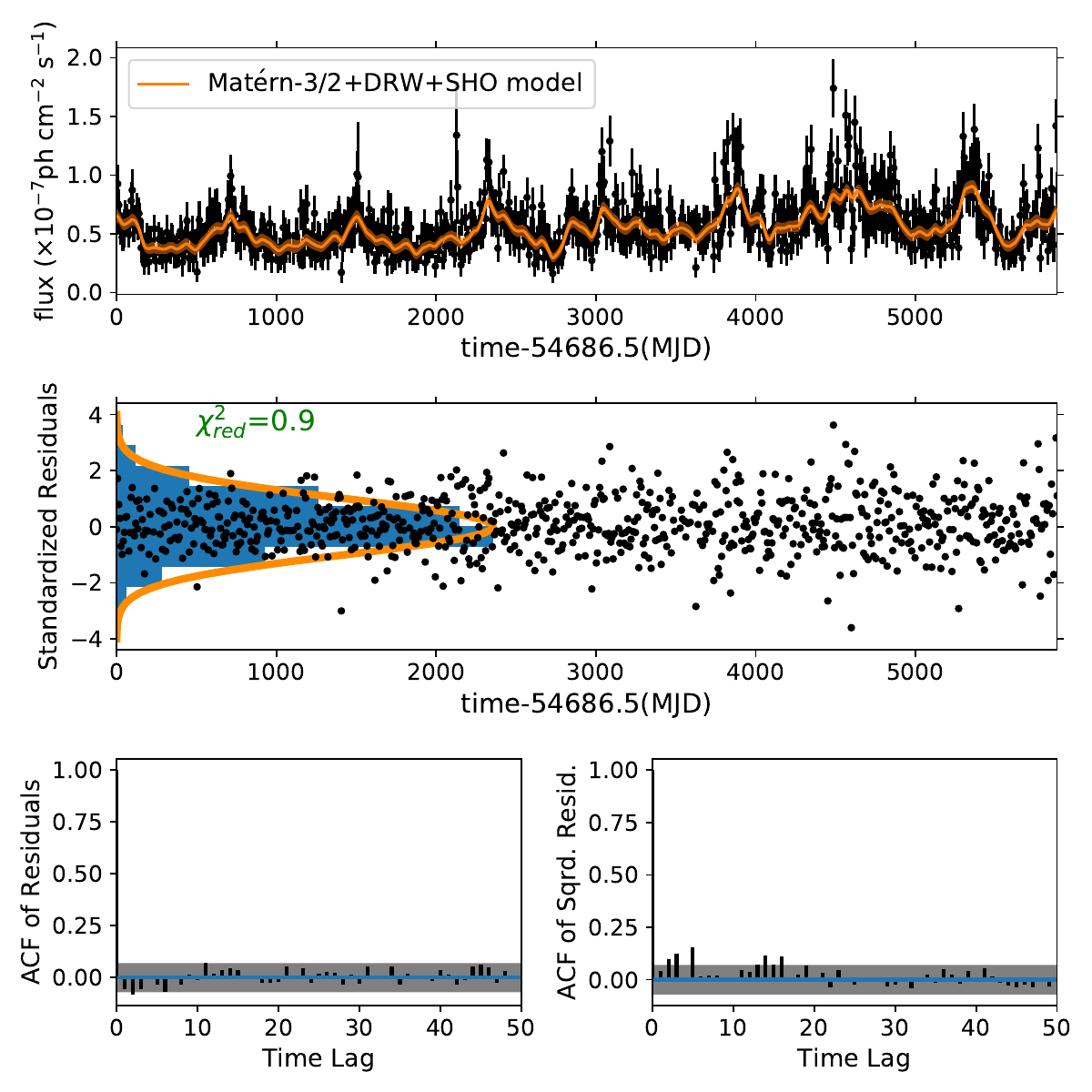}} 
\caption{Fitting results of Mat$\acute{\rm e}$rn$-3/2$+DRW+SHO model for modeling Fermi-LAT light curve of PG 1553+113. The symbols and lines are the same as those in Figure~\ref{fig:DMSfit}.
\label{fig:3_fit}}
\end{figure}

\begin{figure*}
    \centering
    {\includegraphics[width=0.5\linewidth]{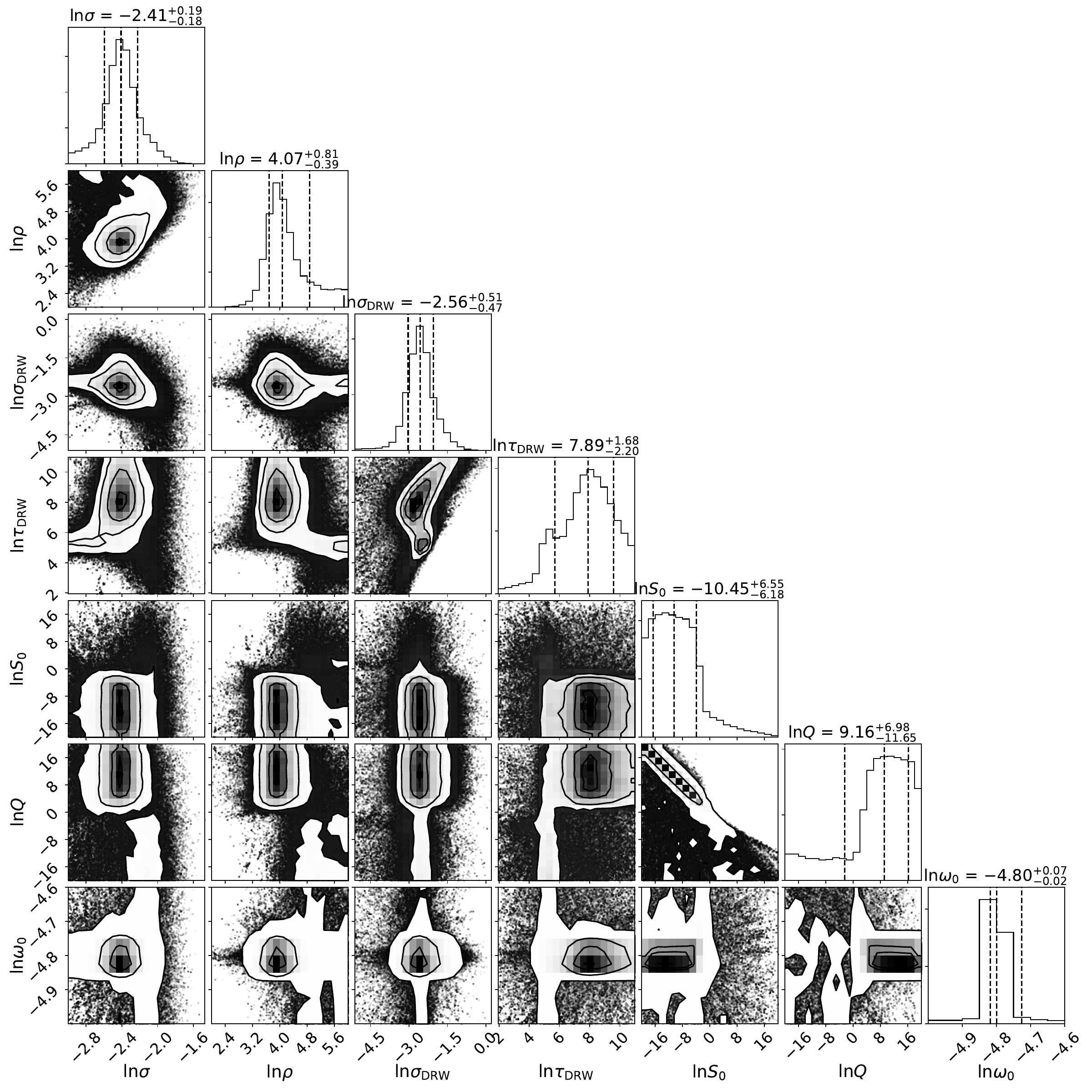}} 
    {\includegraphics[width=0.45\linewidth]{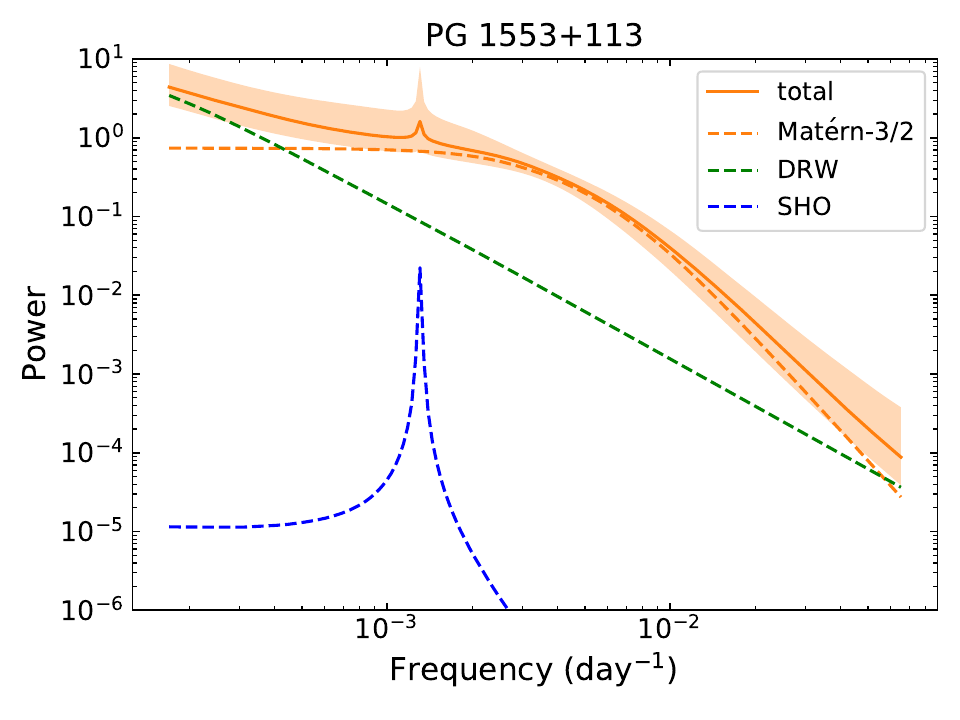}}
\caption{The left panel: posterior parameters distribution of Mat$\acute{\rm e}$rn$-3/2$+DRW+SHO kernel. The symbols and lines are consistent with those in Figure~\ref{fig:DMSparams}. The right panel: the total PSD constructed from the modeling results (the solid orange line and shadow), 
with a dotted orange line, a dotted green line and a dotted blue line representing the Mat$\acute{\rm e}$rn$-3/2$ model, the DRW model, and the SHO model respectively.
\label{fig:3_param+psd}}
\end{figure*}

We then combine DRW, SHO and Mat$\acute{\rm e}$rn$-3/2$ to be 
a new kernel to fit the light curve of PG 1553+113.
According to the goodness-of-fit criteria, 
the fitting is successful (Figure~\ref{fig:3_fit}).
The posterior parameter distribution of the combined model is given in the left panel of Figure~\ref{fig:3_param+psd}.
One can see that $\sigma$, $\rho$ of Mat$\acute{\rm e}$rn$-3/2$ kernel, $\sigma_{\rm DRW}$, $\tau_{\rm DRW}$ of DRW kernel and $\omega_{\rm 0}$ of SHO kernel are robustly constrained.
The parameter $Q$ is more likely to be constrained to large values, with a mean value of 9500.
This suggests that on average the SHO must behave as a nearly natural oscillator.

Indeed, a peak clearly appears in the PSD (right panel of Figure~\ref{fig:3_param+psd}).
The period of the QPO is calculated as:
\begin{equation}
T_{\rm QPO}=\frac{2\pi}{\omega_{\rm 0}\sqrt{1-\xi^{2}}}, \enspace
 \xi=\frac{1}{2Q},\
\label{Tvalue}
\end{equation}
where $Q\gg1$, and $T_{\rm QPO}$ can be approximated as
$2\pi/\omega_{\rm 0}=771\pm 8$ days, which is slightly smaller than that previously reported ($\sim$2.2 years).

The total PSD is generally dominated by the Mat$\acute{\rm e}$rn$-3/2$ component, and the SHO component becomes relevant at $\omega_0\approx0.008\ \rm rad\cdot day^{-1}$.
There is a broken frequency $f_{\rm b}$ in the Mat$\acute{\rm e}$rn$-3/2$ component.
$f_{\rm b}=0.003\pm {0.001}$ $\rm day^{-1}$ corresponds to a timescale $\rho=52^{+23}_{-15}$ days.

The $\rm AIC_{\rm C}=-585$ of this new kernel of Mat$\acute{\rm e}$rn$-3/2$+DRW+SHO is not significantly different from that of the previous two models with $\Delta \rm AIC_{\rm C}\sim3-5$, but its parameter constraints are clearly superior to those of the previous two models. This suggests that the structure in the Fermi-LAT light curve of PG 1553+113 is better described by the Mat$\acute{\rm e}$rn$-3/2$+DRW+SHO model. 

\begin{figure*}
    \centering
    {\includegraphics[width=0.5\linewidth]{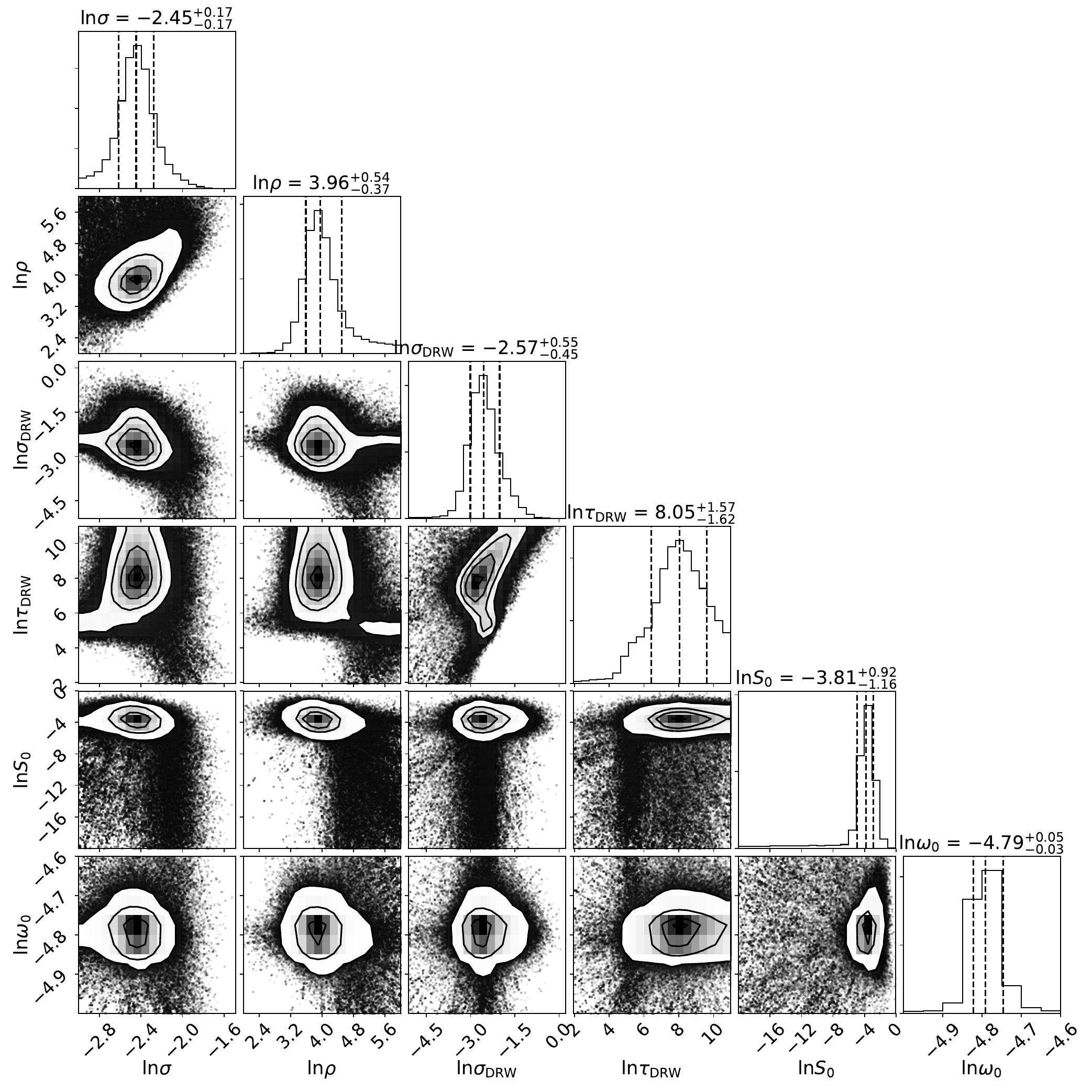}} 
    {\includegraphics[width=0.45\linewidth]{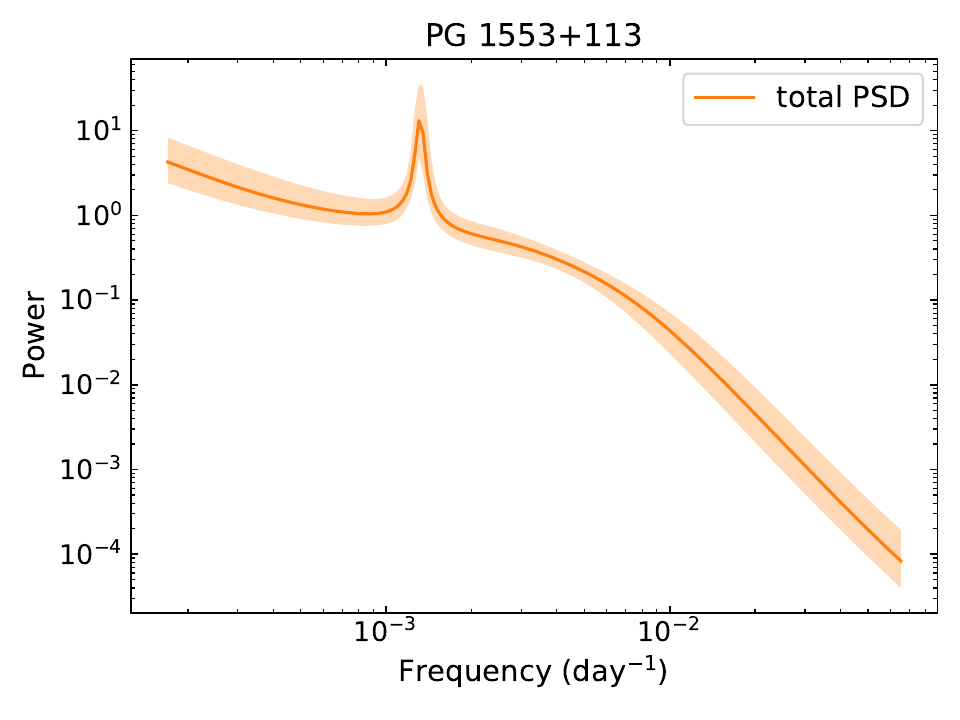}}
\caption{The same as Figure~\ref{fig:3_param+psd}, but with parameter Q fixed to 30.
\label{fig:3_fixQ}}
\end{figure*}

As shown above, without any prior belief about the QPO, 
a QPO signature appears when we select the kernel of Mat$\acute{\rm e}$rn$-3/2$+DRW+SHO.
Now that we have the prior belief about the QPO, we can fix $Q$ to a large value in order to explore the parameter space further.
We fix $Q=30$ (signifying a high-quality oscillator), and then re-fit the light curve.
The results are shown in Figure~\ref{fig:3_fixQ}.
In this case, we prevent the $S_{\rm 0}-Q$ degeneracy, and a more distinct periodic signal appears.

\begin{figure*}
    \centering
    {\includegraphics[width=1\linewidth]{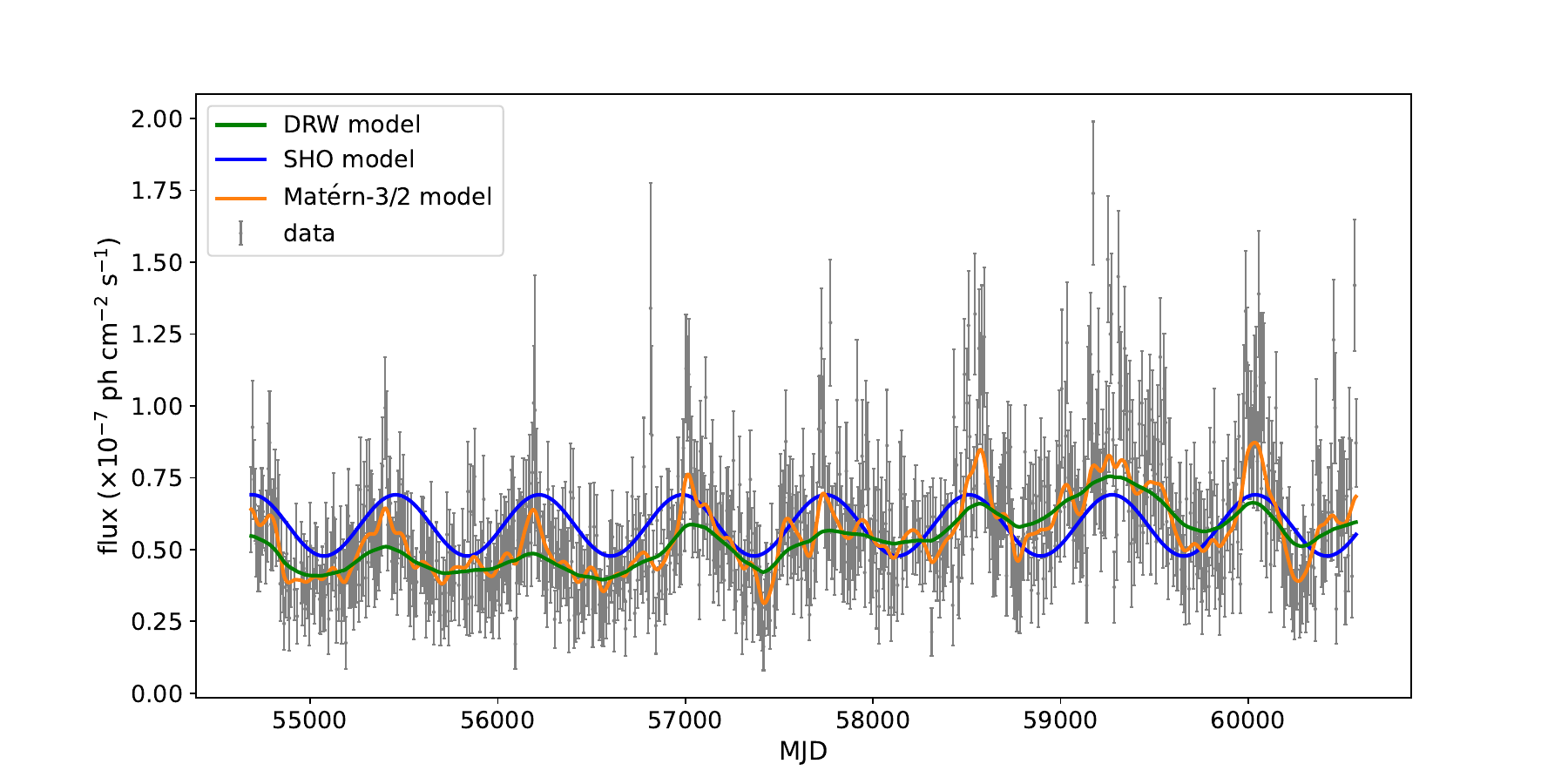}} 
\caption{Predicted light curves of different components in Mat$\acute{\rm e}$rn$-3/2$+DRW+SHO model.
\label{fig:3_pre}}
\end{figure*}

We use three known kernels to reconstruct GP separately, and predict the light curve of each kernel component based on the observational data. 
This allows to separate the total light curve into three components according to the contributions from each kernel, as shown in Figure~\ref{fig:3_pre}.   
The SHO component approaches a high-quality oscillation.
The Mat$\acute{\rm e}$rn$-3/2$ component captures the abrupt changes in the light curve. The DRW component describes the long smooth fluctuations.

\section{Discussion}
High-energy light curves of blazars are generally believed to have significant stochastic components (colored noise), making variability mode exploration and period detection somewhat challenging.
Characterizing the variability mode, especially the stochastic components of light curves accurately is crucial for exploring the underlying physical process of periodic modulation and the stochastic fluctuation as well as the connection between them, i.e., the origin of the variability.

We apply GP method to probe the $\gamma$-ray variability mode of PG 1553+113 using Fermi-LAT data that covers $\sim$ 16 yrs in this work.
The combination of the three kernel functions (DRW, SHO, Mat$\acute{\rm e}$rn$-3/2$), are used to characterize the light curve of PG 1553+113. 
It is found that the Mat$\acute{\rm e}$rn$-3/2$+DRW+SHO combination is the best-fitting model.
In addition to the QPO component with the period of $2.11\pm 0.02$ yr and the common DRW component, a new component of the Mat$\acute{\rm e}$rn$-3/2$ kernel is necessary and dominant.

\begin{figure}
    \centering
    {\includegraphics[width=1\linewidth]{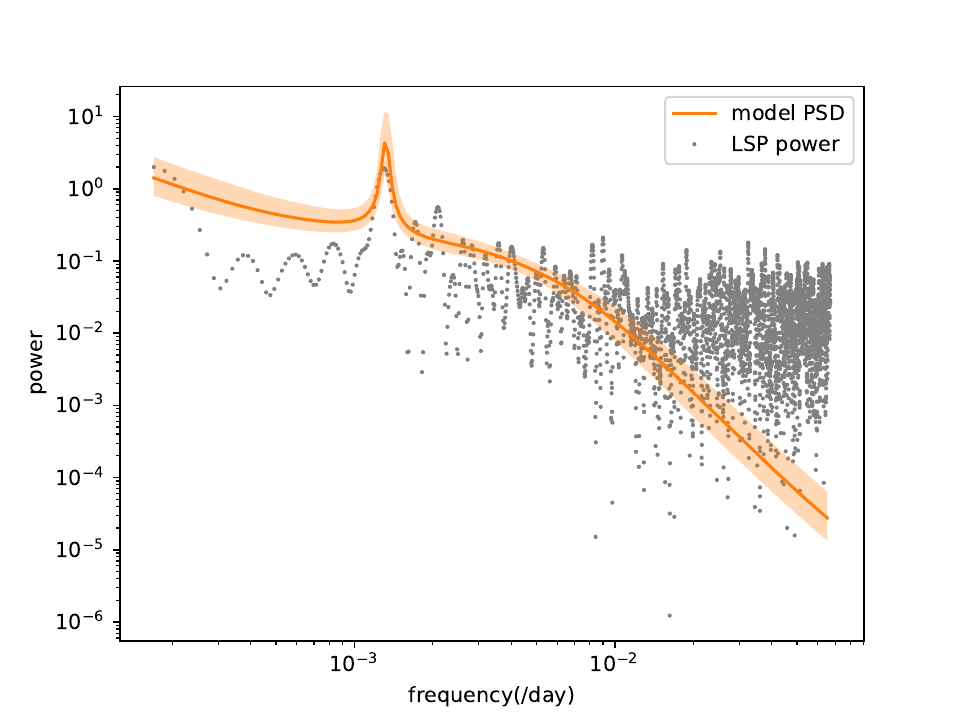}} 
\caption{PSD derived from our GP model (the orange line), and periodogram derived by LSP method (the gray points). 
\label{fig:lsp_psd}}
\end{figure}

Given the light curve data is relatively regular with an average cadence of $\sim$7.5 days, we apply the Lomb-Scargle Periodogram (LSP) method to generate a periodogram and compare it with the PSD derived from our model (see Figure~\ref{fig:lsp_psd}).
A periodic signal of $\sim$771 days is clearly present in the periodogram, which is almost coincides with that in the modeled PSD. From the peak-related frequency to the frequency of 0.01/day, the periodogram and the modeled PSD show a general consistency.
At higher frequencies, the LSP is primarily dominated by the white noise component.
While in our GP analysis, this white noise component represents the measurement uncertainties, which is not shown in the GP PSD.  

We could give a physical explanation for our results in the frame of the supermassive black hole binary (SMBHB) system \citep[e.g.,][]{2017ApJ...836..220C}. 
We only consider the one-jet scenario, namely just one of the supermassive black holes (SMBHs) carrying a relativistic jet.
This jet is a common blazar jet, 
however when it is placed in the binary system, some new and interesting processes emerge.
The periodic component can be naturally understood as being caused by the orbital period of the binary of black holes.
In this situation, as the secondary black hole 
comes close to the primary black hole carrying the jet in its orbit, it exerts a gravitational force on the primary jet that twists and stretches the magnetic field lines within the jet, triggering magnetic reconnection, accelerating particles, and causing periodic flux enhancements \citep{2017ApJ...836..220C,2018ApJ...854...11T}.
As for the DRW component, it aligns with our earlier findings on single black hole systems \citep{2022ApJ...930..157Z,2023ApJ...944..103Z}, where this component represents the common long-term variability resulting from the processes of the accretion disk in the primary black hole system itself \citep{2022ApJ...930..157Z,2023ApJ...944..103Z}.
Unlike our previous results, no effective constraint on the damping timescale was obtained here. This may be due to the energy release triggered by magneto-gravitational stresses in the supermassive binary system continually perturbs the emitting region, preventing it from reaching equilibrium.

More important, our analysis reveals a new component, i.e., the feature that is described by the Mat$\acute{\rm e}$rn$-3/2$ kernel.
It is the dominant component in the data, and mainly describe the abrupt changes in the data. This feature is typically associated with the system undergoing abrupt energy release, such as magnetic reconnection.
Indeed, \cite{2024ApJ...971...26T} showed that 
the PSD of some X-ray bursts of magnetar SGR 1935+2154 is 
similar with that of Mat$\acute{\rm e}$rn$-3/2$ kernel.
Magnetar bursts are usually associated with the magnetic reconnection \citep[][]{2021ApJ...921...92B,2020ApJ...900L..21Y}.
According to these evidence, we suggest that the Mat$\acute{\rm e}$rn$-3/2$ component in PG 1553+113 could be associated with the magnetic reconnection in the jet triggered by the magneto-gravitational stresses in supermassive binaries.
The characteristic timescale $\rho=52^{+23}_{-15}$ days obtained in the Mat$\acute{\rm e}$rn$-3/2$ kernel can be used to derive the size of the region occurring magnetic reconnection \citep[e.g.,][]{2020MNRAS.494.1817D,2022Natur.609..265J}.

\section{Summary}
PG 1553+113 is well known to have periodic variability in the $\gamma$-ray light curve. To gain a deeper understanding of the origin of its $\gamma$-ray variability, we explored its variability mode using GP method.
The combination of Mat$\acute{\rm e}$rn$-3/2$+DRW+SHO demonstrates excellent fitting performance, capturing the complex structures hidden in the light curve.
The Mat$\acute{\rm e}$rn$-3/2$ component is a previously unrecognized pattern in active galactic nuclei variability studies, which is typically associated with the system undergoing abrupt energy release, such as magnetic reconnection. 
These findings can be consistently interpreted within the SMBHB framework suggested by \cite{2017ApJ...836..220C}.
In this scenario, the magnetic reconnection in blazar jet triggered by the magneto-gravitational stresses not only produces the QPO signature, but also the aperiodic abrupt changes in the light curve.

\section*{Data Availability}
The $\gamma$-ray data used in this work are publicly available and can be found at the Fermi-LAT light curve repository \url{https://fermi.gsfc.nasa.gov/ssc/data/access/lat/LightCurveRepository/}. The celerite code can be referenced from \url{https://celerite.readthedocs.io/en/stable/}.
The data and Python code for modeling light curves are available on GitHub: \url{https://github.com/Hyun-zhang/Astronomical-application-of-GP-Celerite.git}.

\section*{acknowledgments}
D.H.Y acknowledges funding support from the National Natural Science Foundation of China (NSFC) under grant No. 12393852. We thank the support from the Postdoctoral Fellowship Program of CPSF under Grant Number GZB20230618.

{\it Facility:} Fermi(LAT)

{\it Software:} Fermitools-conda, celerite \citep{2017AJ....154..220F}, emcee \citep{2013PASP..125..306F}, NumPy \citep{2020NumPy-Array}, Matplotlib \citep{2007CSE.....9...90H}, Astropy \citep{2013A&A...558A..33A,2018AJ....156..123A}, SciPy \citep{2020SciPy-NMeth}.

\bibliography{main}

\begin{thebibliography}{}
\makeatletter
\relax
\def\mn@urlcharsother{\let\do\@makeother \do\$\do\&\do\#\do\^\do\_\do\%\do\~}
\def\mn@doi{\begingroup\mn@urlcharsother \@ifnextchar [ {\mn@doi@} {\mn@doi@[]}}
\def\mn@doi@[#1]#2{\def\@tempa{#1}\ifx\@tempa\@empty \href {http://dx.doi.org/#2} {doi:#2}\else \href {http://dx.doi.org/#2} {#1}\fi \endgroup}
\def\mn@eprint#1#2{\mn@eprint@#1:#2::\@nil}
\def\mn@eprint@arXiv#1{\href {http://arxiv.org/abs/#1} {{\tt arXiv:#1}}}
\def\mn@eprint@dblp#1{\href {http://dblp.uni-trier.de/rec/bibtex/#1.xml} {dblp:#1}}
\def\mn@eprint@#1:#2:#3:#4\@nil{\def\@tempa {#1}\def\@tempb {#2}\def\@tempc {#3}\ifx \@tempc \@empty \let \@tempc \@tempb \let \@tempb \@tempa \fi \ifx \@tempb \@empty \def\@tempb {arXiv}\fi \@ifundefined {mn@eprint@\@tempb}{\@tempb:\@tempc}{\expandafter \expandafter \csname mn@eprint@\@tempb\endcsname \expandafter{\@tempc}}}

\bibitem[\protect\citeauthoryear{{Ackermann} et~al.,}{{Ackermann} et~al.}{2015}]{2015ApJ...813L..41A}
{Ackermann} M.,  et~al., 2015, \mn@doi [\apjl] {10.1088/2041-8205/813/2/L41}, \href {https://ui.adsabs.harvard.edu/abs/2015ApJ...813L..41A} {813, L41}

\bibitem[\protect\citeauthoryear{{Aigrain} \& {Foreman-Mackey}}{{Aigrain} \& {Foreman-Mackey}}{2023}]{2023ARA&A..61..329A}
{Aigrain} S.,  {Foreman-Mackey} D.,  2023, \mn@doi [\araa] {10.1146/annurev-astro-052920-103508}, \href {https://ui.adsabs.harvard.edu/abs/2023ARA&A..61..329A} {61, 329}

\bibitem[\protect\citeauthoryear{{Astropy Collaboration} et~al.,}{{Astropy Collaboration} et~al.}{2013}]{2013A&A...558A..33A}
{Astropy Collaboration} et~al., 2013, \mn@doi [\aap] {10.1051/0004-6361/201322068}, \href {https://ui.adsabs.harvard.edu/abs/2013A&A...558A..33A} {558, A33}

\bibitem[\protect\citeauthoryear{{Astropy Collaboration} et~al.,}{{Astropy Collaboration} et~al.}{2018}]{2018AJ....156..123A}
{Astropy Collaboration} et~al., 2018, \mn@doi [\aj] {10.3847/1538-3881/aabc4f}, \href {https://ui.adsabs.harvard.edu/abs/2018AJ....156..123A} {156, 123}

\bibitem[\protect\citeauthoryear{{Beloborodov}}{{Beloborodov}}{2021}]{2021ApJ...921...92B}
{Beloborodov} A.~M.,  2021, \mn@doi [\apj] {10.3847/1538-4357/ac17e7}, \href {https://ui.adsabs.harvard.edu/abs/2021ApJ...921...92B} {921, 92}

\bibitem[\protect\citeauthoryear{{Britzen} et~al.,}{{Britzen} et~al.}{2018}]{2018MNRAS.478.3199B}
{Britzen} S.,  et~al., 2018, \mn@doi [\mnras] {10.1093/mnras/sty1026}, \href {https://ui.adsabs.harvard.edu/abs/2018MNRAS.478.3199B} {478, 3199}

\bibitem[\protect\citeauthoryear{{Burke} et~al.,}{{Burke} et~al.}{2021}]{2021Sci...373..789B}
{Burke} C.~J.,  et~al., 2021, \mn@doi [Science] {10.1126/science.abg9933}, \href {https://ui.adsabs.harvard.edu/abs/2021Sci...373..789B} {373, 789}

\bibitem[\protect\citeauthoryear{{Cavaliere}, {Tavani}  \& {Vittorini}}{{Cavaliere} et~al.}{2017}]{2017ApJ...836..220C}
{Cavaliere} A.,  {Tavani} M.,   {Vittorini} V.,  2017, \mn@doi [\apj] {10.3847/1538-4357/836/2/220}, \href {https://ui.adsabs.harvard.edu/abs/2017ApJ...836..220C} {836, 220}

\bibitem[\protect\citeauthoryear{{Cavaliere}, {Tavani}, {Munar-Adrover}  \& {Argan}}{{Cavaliere} et~al.}{2019}]{2019ApJ...875L..22C}
{Cavaliere} A.,  {Tavani} M.,  {Munar-Adrover} P.,   {Argan} A.,  2019, \mn@doi [\apjl] {10.3847/2041-8213/ab0e88}, \href {https://ui.adsabs.harvard.edu/abs/2019ApJ...875L..22C} {875, L22}

\bibitem[\protect\citeauthoryear{{Covino}, {Sandrinelli}  \& {Treves}}{{Covino} et~al.}{2019}]{2019MNRAS.482.1270C}
{Covino} S.,  {Sandrinelli} A.,   {Treves} A.,  2019, \mn@doi [\mnras] {10.1093/mnras/sty2720}, \href {https://ui.adsabs.harvard.edu/abs/2019MNRAS.482.1270C} {482, 1270}

\bibitem[\protect\citeauthoryear{{Covino}, {Landoni}, {Sandrinelli}  \& {Treves}}{{Covino} et~al.}{2020}]{2020ApJ...895..122C}
{Covino} S.,  {Landoni} M.,  {Sandrinelli} A.,   {Treves} A.,  2020, \mn@doi [\apj] {10.3847/1538-4357/ab8bd4}, \href {https://ui.adsabs.harvard.edu/abs/2020ApJ...895..122C} {895, 122}

\bibitem[\protect\citeauthoryear{{Dong}, {Zhang}  \& {Giannios}}{{Dong} et~al.}{2020}]{2020MNRAS.494.1817D}
{Dong} L.,  {Zhang} H.,   {Giannios} D.,  2020, \mn@doi [\mnras] {10.1093/mnras/staa773}, \href {https://ui.adsabs.harvard.edu/abs/2020MNRAS.494.1817D} {494, 1817}

\bibitem[\protect\citeauthoryear{{Foreman-Mackey}, {Hogg}, {Lang}  \& {Goodman}}{{Foreman-Mackey} et~al.}{2013}]{2013PASP..125..306F}
{Foreman-Mackey} D.,  {Hogg} D.~W.,  {Lang} D.,   {Goodman} J.,  2013, \mn@doi [\pasp] {10.1086/670067}, \href {https://ui.adsabs.harvard.edu/abs/2013PASP..125..306F} {125, 306}

\bibitem[\protect\citeauthoryear{{Foreman-Mackey}, {Agol}, {Ambikasaran}  \& {Angus}}{{Foreman-Mackey} et~al.}{2017}]{2017AJ....154..220F}
{Foreman-Mackey} D.,  {Agol} E.,  {Ambikasaran} S.,   {Angus} R.,  2017, \mn@doi [\aj] {10.3847/1538-3881/aa9332}, \href {https://ui.adsabs.harvard.edu/abs/2017AJ....154..220F} {154, 220}

\bibitem[\protect\citeauthoryear{{Gao}, {Lu}, {Qin}, {Gong}, {Yu}, {Li}  \& {Yi}}{{Gao} et~al.}{2023}]{2023ApJ...945..146G}
{Gao} Q.-G.,  {Lu} F.-W.,  {Qin} L.-h.,  {Gong} Y.-L.,  {Yu} G.-m.,  {Li} H.-z.,   {Yi} T.-f.,  2023, \mn@doi [\apj] {10.3847/1538-4357/acbe3e}, \href {https://ui.adsabs.harvard.edu/abs/2023ApJ...945..146G} {945, 146}

\bibitem[\protect\citeauthoryear{Harris et~al.,}{Harris et~al.}{2020}]{2020NumPy-Array}
Harris C.~R.,  et~al., 2020, \mn@doi [Nature] {10.1038/s41586-020-2649-2}, 585, 357–362

\bibitem[\protect\citeauthoryear{{Huang}, {Yin}, {Hu}, {Chen}, {Jiang}, {Alexeeva}  \& {Wang}}{{Huang} et~al.}{2021}]{2021ApJ...922..222H}
{Huang} S.,  {Yin} H.,  {Hu} S.,  {Chen} X.,  {Jiang} Y.,  {Alexeeva} S.,   {Wang} Y.,  2021, \mn@doi [\apj] {10.3847/1538-4357/ac2d98}, \href {https://ui.adsabs.harvard.edu/abs/2021ApJ...922..222H} {922, 222}

\bibitem[\protect\citeauthoryear{{Hunter}}{{Hunter}}{2007}]{2007CSE.....9...90H}
{Hunter} J.~D.,  2007, \mn@doi [Computing in Science and Engineering] {10.1109/MCSE.2007.55}, \href {https://ui.adsabs.harvard.edu/abs/2007CSE.....9...90H} {9, 90}

\bibitem[\protect\citeauthoryear{{Jorstad} et~al.,}{{Jorstad} et~al.}{2022}]{2022Natur.609..265J}
{Jorstad} S.~G.,  et~al., 2022, \mn@doi [\nat] {10.1038/s41586-022-05038-9}, \href {https://ui.adsabs.harvard.edu/abs/2022Natur.609..265J} {609, 265}

\bibitem[\protect\citeauthoryear{{Moreno}, {Vogeley}, {Richards}  \& {Yu}}{{Moreno} et~al.}{2019}]{2019PASP..131f3001M}
{Moreno} J.,  {Vogeley} M.~S.,  {Richards} G.~T.,   {Yu} W.,  2019, \mn@doi [\pasp] {10.1088/1538-3873/ab1597}, \href {https://ui.adsabs.harvard.edu/abs/2019PASP..131f3001M} {131, 063001}

\bibitem[\protect\citeauthoryear{{O'Sullivan} \& {Aigrain}}{{O'Sullivan} \& {Aigrain}}{2024}]{2024MNRAS.531.4181O}
{O'Sullivan} N.~K.,  {Aigrain} S.,  2024, \mn@doi [\mnras] {10.1093/mnras/stae1059}, \href {https://ui.adsabs.harvard.edu/abs/2024MNRAS.531.4181O} {531, 4181}

\bibitem[\protect\citeauthoryear{{Rasmussen} \& {Williams}}{{Rasmussen} \& {Williams}}{2006}]{2006gpml.book.....R}
{Rasmussen} C.~E.,  {Williams} C. K.~I.,  2006, {Gaussian Processes for Machine Learning}

\bibitem[\protect\citeauthoryear{{Ryan}, {Siemiginowska}, {Sobolewska}  \& {Grindlay}}{{Ryan} et~al.}{2019}]{2019ApJ...885...12R}
{Ryan} J.~L.,  {Siemiginowska} A.,  {Sobolewska} M.~A.,   {Grindlay} J.,  2019, \mn@doi [\apj] {10.3847/1538-4357/ab426a}, \href {https://ui.adsabs.harvard.edu/abs/2019ApJ...885...12R} {885, 12}

\bibitem[\protect\citeauthoryear{{Sobacchi}, {Sormani}  \& {Stamerra}}{{Sobacchi} et~al.}{2017}]{2017MNRAS.465..161S}
{Sobacchi} E.,  {Sormani} M.~C.,   {Stamerra} A.,  2017, \mn@doi [\mnras] {10.1093/mnras/stw2684}, \href {https://ui.adsabs.harvard.edu/abs/2017MNRAS.465..161S} {465, 161}

\bibitem[\protect\citeauthoryear{{Tang} et~al.,}{{Tang} et~al.}{2024}]{2024ApJ...971...26T}
{Tang} R.,  et~al., 2024, \mn@doi [\apj] {10.3847/1538-4357/ad5a03}, \href {https://ui.adsabs.harvard.edu/abs/2024ApJ...971...26T} {971, 26}

\bibitem[\protect\citeauthoryear{{Tavani}, {Cavaliere}, {Munar-Adrover}  \& {Argan}}{{Tavani} et~al.}{2018}]{2018ApJ...854...11T}
{Tavani} M.,  {Cavaliere} A.,  {Munar-Adrover} P.,   {Argan} A.,  2018, \mn@doi [\apj] {10.3847/1538-4357/aaa3f4}, \href {https://ui.adsabs.harvard.edu/abs/2018ApJ...854...11T} {854, 11}

\bibitem[\protect\citeauthoryear{{Vaughan}, {Uttley}, {Markowitz}, {Huppenkothen}, {Middleton}, {Alston}, {Scargle}  \& {Farr}}{{Vaughan} et~al.}{2016}]{2016MNRAS.461.3145V}
{Vaughan} S.,  {Uttley} P.,  {Markowitz} A.~G.,  {Huppenkothen} D.,  {Middleton} M.~J.,  {Alston} W.~N.,  {Scargle} J.~D.,   {Farr} W.~M.,  2016, \mn@doi [\mnras] {10.1093/mnras/stw1412}, \href {https://ui.adsabs.harvard.edu/abs/2016MNRAS.461.3145V} {461, 3145}

\bibitem[\protect\citeauthoryear{{Villata}, {Raiteri}, {Sillanpaa}  \& {Takalo}}{{Villata} et~al.}{1998}]{1998MNRAS.293L..13V}
{Villata} M.,  {Raiteri} C.~M.,  {Sillanpaa} A.,   {Takalo} L.~O.,  1998, \mn@doi [\mnras] {10.1046/j.1365-8711.1998.01244.x}, \href {https://ui.adsabs.harvard.edu/abs/1998MNRAS.293L..13V} {293, L13}

\bibitem[\protect\citeauthoryear{Virtanen et~al.,}{Virtanen et~al.}{2020}]{2020SciPy-NMeth}
Virtanen P.,  et~al., 2020, \mn@doi [Nature Methods] {10.1038/s41592-019-0686-2}, \href {https://rdcu.be/b08Wh} {17, 261}

\bibitem[\protect\citeauthoryear{{Yan}, {Zhou}, {Zhang}, {Zhu}  \& {Wang}}{{Yan} et~al.}{2018}]{2018ApJ...867...53Y}
{Yan} D.,  {Zhou} J.,  {Zhang} P.,  {Zhu} Q.,   {Wang} J.,  2018, \mn@doi [\apj] {10.3847/1538-4357/aae48a}, \href {https://ui.adsabs.harvard.edu/abs/2018ApJ...867...53Y} {867, 53}

\bibitem[\protect\citeauthoryear{{Yang}, {Yan}, {Zhang}, {Dai}  \& {Zhang}}{{Yang} et~al.}{2021}]{2021ApJ...907..105Y}
{Yang} S.,  {Yan} D.,  {Zhang} P.,  {Dai} B.,   {Zhang} L.,  2021, \mn@doi [\apj] {10.3847/1538-4357/abcbff}, \href {https://ui.adsabs.harvard.edu/abs/2021ApJ...907..105Y} {907, 105}

\bibitem[\protect\citeauthoryear{{Yuan}, {Beloborodov}, {Chen}  \& {Levin}}{{Yuan} et~al.}{2020}]{2020ApJ...900L..21Y}
{Yuan} Y.,  {Beloborodov} A.~M.,  {Chen} A.~Y.,   {Levin} Y.,  2020, \mn@doi [\apjl] {10.3847/2041-8213/abafa8}, \href {https://ui.adsabs.harvard.edu/abs/2020ApJ...900L..21Y} {900, L21}

\bibitem[\protect\citeauthoryear{{Zhang}, {Yan}, {Zhang}, {Yang}  \& {Zhang}}{{Zhang} et~al.}{2021}]{2021ApJ...919...58Z}
{Zhang} H.,  {Yan} D.,  {Zhang} P.,  {Yang} S.,   {Zhang} L.,  2021, \mn@doi [\apj] {10.3847/1538-4357/ac0cf0}, \href {https://ui.adsabs.harvard.edu/abs/2021ApJ...919...58Z} {919, 58}

\bibitem[\protect\citeauthoryear{{Zhang}, {Yan}  \& {Zhang}}{{Zhang} et~al.}{2022}]{2022ApJ...930..157Z}
{Zhang} H.,  {Yan} D.,   {Zhang} L.,  2022, \mn@doi [\apj] {10.3847/1538-4357/ac679e}, \href {https://ui.adsabs.harvard.edu/abs/2022ApJ...930..157Z} {930, 157}

\bibitem[\protect\citeauthoryear{{Zhang}, {Yan}  \& {Zhang}}{{Zhang} et~al.}{2023}]{2023ApJ...944..103Z}
{Zhang} H.,  {Yan} D.,   {Zhang} L.,  2023, \mn@doi [\apj] {10.3847/1538-4357/acafe5}, \href {https://ui.adsabs.harvard.edu/abs/2023ApJ...944..103Z} {944, 103}

\makeatother
\end{thebibliography}
\bibliographystyle{mnras}

\end{document}